\documentclass[aps,prl,twocolumn,superscriptaddress]{revtex4-1}

\usepackage{amsmath,amssymb,mathtools}
\usepackage{mathrsfs}
\usepackage{bbold}
\usepackage{epsfig}
\usepackage{graphicx,color}
\usepackage[colorlinks=true,linkcolor=blue,citecolor=blue]{hyperref}

\begin{document}
	
\title{The diffractive saturable loss mechanism in Kerr-lens mode-locked lasers: direct observation and simulation}
	
\author{Idan Parshani}
\thanks{These two authors contributed equally.}
\author{Leon Bello}
\thanks{These two authors contributed equally.}
\author{Mallachi-Elia Meller}
\author{Avi Pe'er}

\affiliation{Department of Physics and BINA Center of Nanotechnology, Bar-Ilan University, 5290002 Ramat-Gan, Israel}




\date{\today}

\begin{abstract}
Passive mode-locking critically relies on a saturable loss mechanism to form ultrashort pulses. 
However, in Kerr-lens mode-locking (KLM), no actual absorption takes place, but rather losses appear due to diffraction, and actual light must escape the cavity. The Kerr-lens effect works to generate through diffraction an effective instantaneous saturable absorber that delicately depends on the interplay between the spatial and temporal profiles of the pulse. Despite the importance of KLM as a technique for generating ultrafast pulses and the fundamental role of the diffraction losses in its operation, these losses were never directly observed. Here, we measure the light that leaks out due to diffraction losses in a hard-aperture Kerr-lens mode-locked Ti:Sapphire laser, and compare the measured results with a numerical theory that explicitly calculates the spatio-temporal behavior of the pulse.
\end{abstract}

\maketitle

A mode-locked laser oscillates on many longitudinal modes together in a synchronized manner such that the modes interfere constructively to create a short pulse in time.  
Mode-locking is achieved by incorporating in the cavity a saturable loss mechanism, i.e. a saturable absorber (SA), whose loss is reduced at high intensities \cite{Ippen1994, Kaertner2005}. The intuition is that the laser continuously optimizes to find the most efficient mode of operation, which in the case of a SA, corresponds to mode-locked operation, where the high peak-intensity of the pulses mitigates the absorption losses. Saturable absorbers can be roughly divided into fast and slow SAs, and their dynamics differ considerably \cite{Kurtner1998}. In this work, we focus on Kerr-lens mode-locking, which is an extreme example of a fast (practically instantaneous) SA. In fast saturable absorbers, the losses directly vary with the power in time, indicating that the SA is more absorptive at the leading and trailing edges of the pulse, which leaves more net gain to the center part (peak) of pulse. This works to shorten the pulse, until pulse-widening effects like dispersion and finite gain bandwidth become substantial and the pulse reaches a steady-state. Understanding the dynamics that lead to minimum-loss solutions is an important fundamental problem in the study of laser physics in general and mode-locking in particular \cite{Wright2020}.
\begin{figure}[htbp]
	\centering 
	\fbox{\includegraphics[width=0.75\linewidth]{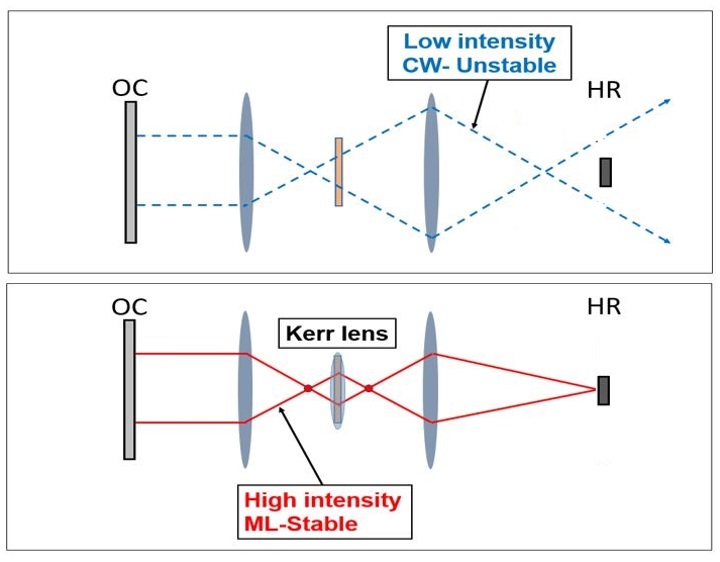}}
	\caption{KLM Basic mechanism: \textbf{(top)} In the common KLM configuration, the cavity is often operated at the point where CW operation is unstable and the diffraction losses are very sensitive to changes in power. \textbf{(bottom)} For high-intensity pulses, the non-linear medium acts as an effective non-linear lens, making the cavity stable under mode-locked operation and mitigating the losses.}
	\label{fig:configuration}
\end{figure}
Mode-locking is a very important example of mode-competition, where through the dynamics of gain, loss and non-linearities a laser finds the most efficient mode of operation, rendering it as one of the most powerful  optimization devices. Using the principles of mode competition, lasers have been used as real-time optical solvers \cite{Nixon2013} and even simulators for spin-networks \cite{Marandi2014, Nixon2013a}. In mode locking, the additional non-linearity of the loss, which acts in both space and time, changes the most-efficient mode from a single frequency continuous-wave mode to a single highly localized temporal mode, i.e. a pulse. 

Kerr-lens mode-locking (KLM) is the major technique for producing ultrafast laser pulses, enabling the generation of the shortest state-of-the-art pulses in Ti:Sapphire lasers \cite{Morgner1999}. We present a detailed experimental study of the spatio-temporal dynamics of a KLM Ti:Sapphire laser. Despite it being very well known that KLM is a spatio-temporal effect that involves losses exclusively due to light diffracting outside of the cavity, it was never directly observed in hard-aperture KLM.
 
KLM works by creating an effective saturable absorber through the non-linear index of refraction of the Ti:Sapphire, or any other non-linear media inside the cavity \cite{meller_2017, Yefet2013}. The intensity dependent refractive index $n(I) = n_0 + n_2 I$, together with the spatial profile of the beam generates an effective non-linear lens inside the cavity, whose focal length depends on the instantaneous beam radius $w(t)$ and power $P(t)$ at the Kerr medium according to $\frac{1}{f_{nl}} = \kappa_{nl} \frac{P(t)}{w^4(t)}$, where $\kappa_{nl}$ is the non-linear response of the Kerr medium. The saturable absorber operation in mode-locking involves a number of mechanisms. In soft-aperture KLM, the losses are due to the mismatch between the lasing and pump modes in the gain medium. In hard-aperture KLM, a physical element, acting as an aperture, is present in the cavity and causes diffraction losses for diverging beams. In our experiment, the hard aperture is virtual, since light at the edges of the pulse, which does not experience the stabilizing Kerr lens effect, travels in an unstable cavity and is bound to leave the beam no matter what (with an aperture or not). The additional instantaneous non-linear lensing can mitigate these losses for high intensities, acting as an effective \emph{ultrafast} saturable absorber \cite{yefet_2013}. Other non-linear effects may also play an important role, but in this work we focus on the Kerr-induced diffraction losses. The virtually instantaneous nature of the Kerr-lens is key to achieve extremely short pulses. While the core function of the Kerr-lens is the same as a saturable absorber, clearly no actual absorption takes place. KLM is an inherently spatial effect and losses are entirely due to diffraction, indicating that in hard-aperture KLM light must actively leak outside the cavity. In fact, at the advent of KLM, the role of the Kerr-lens effect and diffraction losses was not understood \cite{Spence1991, Salin1991}, and in older literature, KLM is called self mode-locking, due to the apparent lack of a visible absorber. Even today, the fully coupled spatiotemporal dynamics of KLM is not completely understood or modeled, and the effectiveness of the saturable absorber is in many cases approximated with a description in terms of Ginzburg-Landau equations \cite{haus_2000}, where the spatial nature of KLM is neglected, or analyzed in isolation from the temporal behavior of the pulse in a cavity stability analysis \cite{magni_1987, krausz_1992}. 

\begin{figure}[htbp]
	\centering
	\fbox{\includegraphics[width=0.95\linewidth]{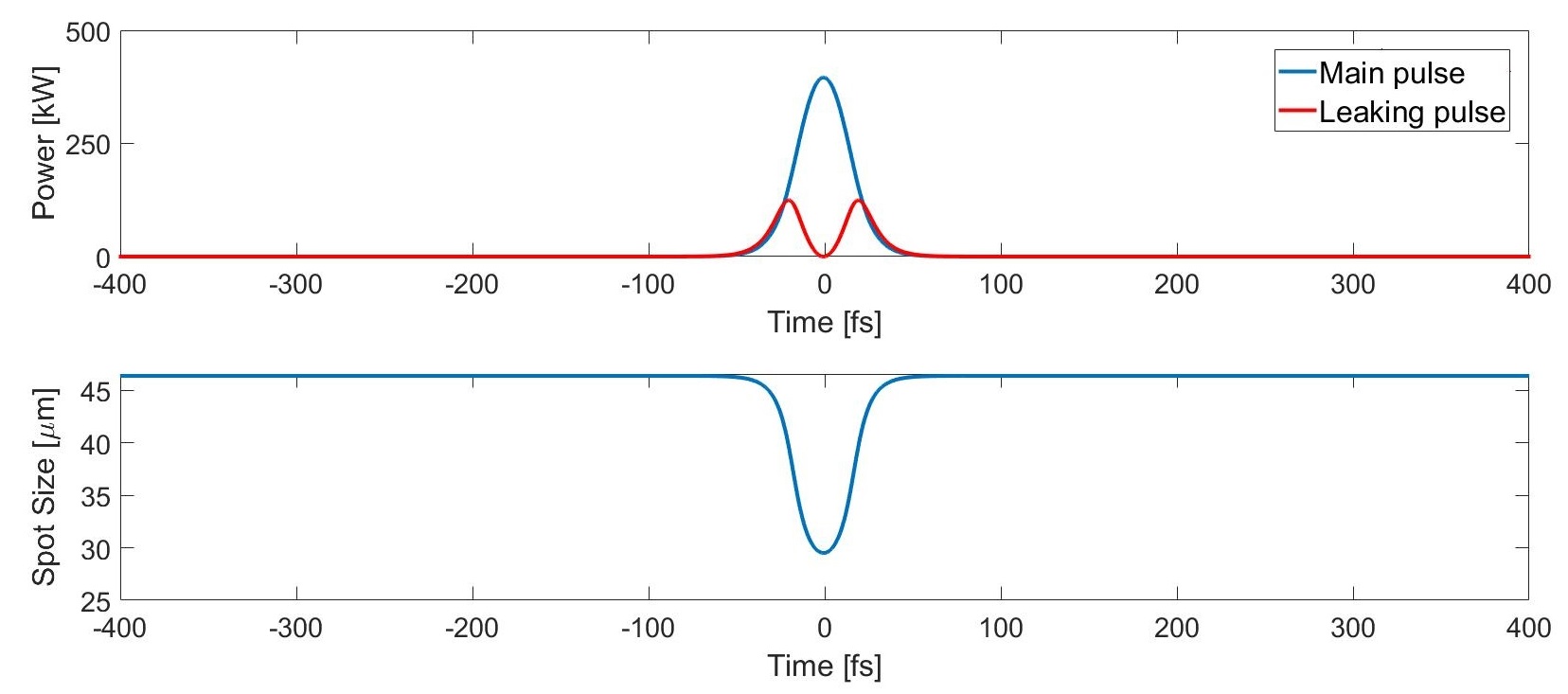}}
	\caption{\textbf{Top:} Calculated temporal profile of the main pulse (blue) and of the leaking light (red). Since the light leaks at the edges of the main pulse, the leaking light appears as a double pulse. \textbf{Bottom:} Calculated instantaneous beam radius $w(t)$ on the end mirror, showing the strong time dependence of the spot-size.}
	\label{fig:temporal_profile}
\end{figure} 

Since light in KLM is only diffracted out, it can be directly measured to reveal the spatio-temporal behavior of the pulse. In this work, we calculate, simulate and directly observe the diffraction losses in KLM. We use a numerical approach to simulate the KLM dynamics both in time and space, taking into account the interplay between the spatial and temporal parts, which is usually neglected in the analysis of KLM. 

\begin{figure}[ht!]
	\centering
	\fbox{\includegraphics[width=0.9\linewidth]{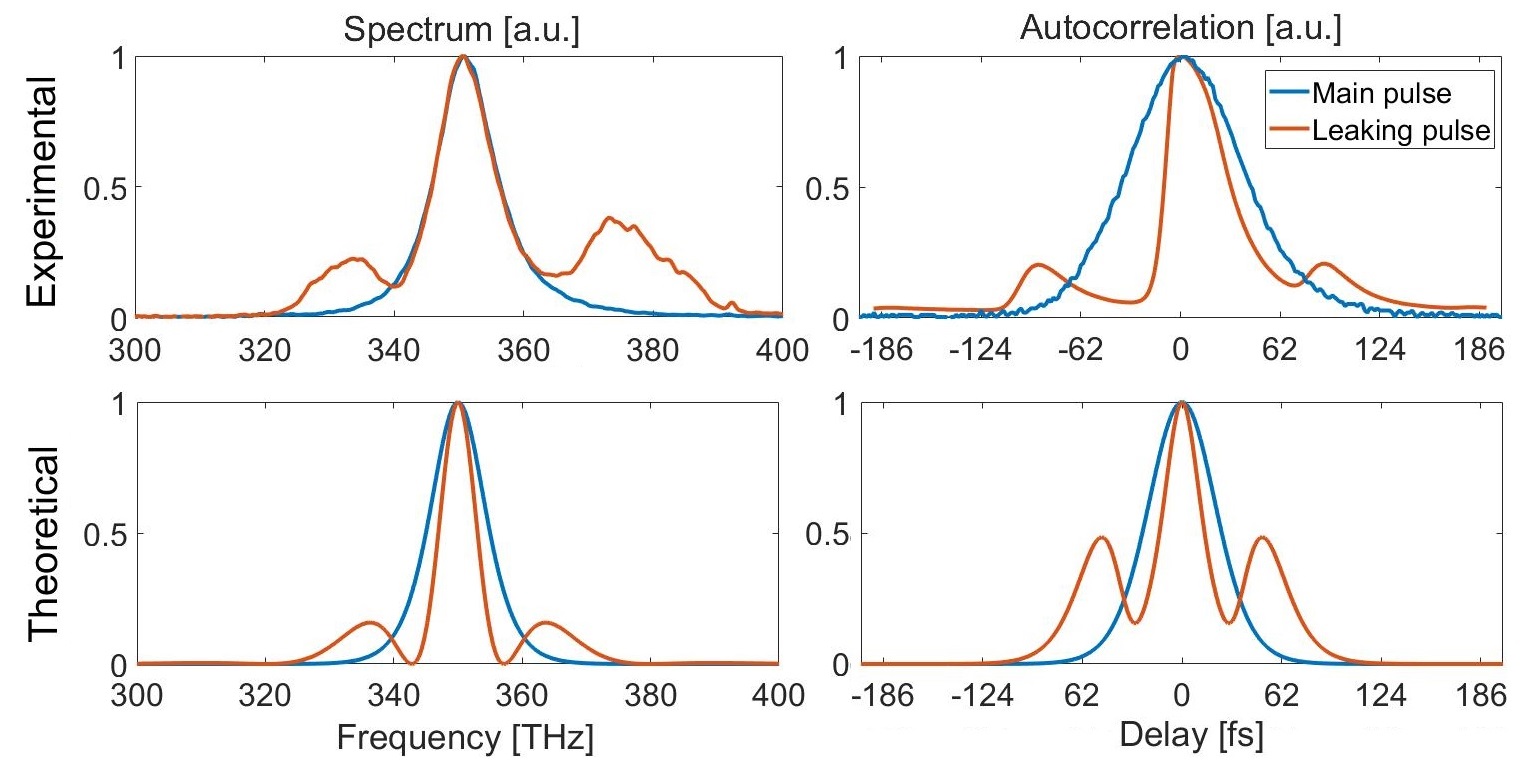}}
	\caption{Theory \textbf{(left)} vs. measured \textbf{(right)} light leakage out of the cavity in both time (top - intensity auto-correlation) and frequency (bottom - spectrum). The experimental results show very good agreement with the simulated one (left).}
	\label{fig:autocorrelation}
\end{figure}

Our analysis focuses on the hard-aperture effective SA due to diffraction losses. Other important effects are also present in KLM, such as gain-laser coupling and gain saturation, but in some regimes the diffraction losses are a dominant and sufficient effect, and these other effects can be neglected. Gain modulation due to the laser-pump coupling is incorporated in our analysis as gain depletion due to laser-pump interaction in the Ti:Sapphire crystal. The only approximation taken was to assume the pulse is much shorter than the pumping time scale, indicating that gain recovery is assumed only between pulses, but neglected within the ultrashort duration of the laser pulse. This assumption is well validated for femtosecond pulses in a cw-pumped Ti:Sapphire oscillator, as we have in the experiment.
Our reduced model successfully captures the dynamics of KLM, and allows us to go beyond the steady-state analysis to observe the real-time dynamical behavior of the oscillation. It yields the instantaneous relation between the pulse power and the beam radius, as well as the temporal profile of the leaking light and the pulse, as shown in figures \ref{fig:temporal_profile} and \ref{fig:autocorrelation}. 

In order to make the calculation tractable, we approximate the spatial mode of the laser to be a single transverse Gaussian $TEM_{00}$ mode, whose waist is intensity dependent and varies in time with the pulse through the Kerr-lens effect. Under the Gaussian approximation, the spatial part can be fully represented by a time-dependent complex beam parameter $q_{n}(t)$ that is propagated through the cavity by standard linear ABCD propagation with a propagation matrix that depends on time through the intensity dependent Kerr-lens for each moment in time \cite{Siegman_ABCD}. 
Our analysis incorporates both the temporal and spatial evolution of the intra-cavity field. Our simulation makes no note of the carrier frequency, which is completely arbitrary, and simulates only the envelopes of the pulse. The temporal envelope of the field $E_{n}(t)$ is divided into two time-scales - the fast time within a round-trip $t$ and \textcolor{black}{the slower time scale between round-trips $\tau_{n}$, expressed in multiples of the round-trip time, and is incorporated by repeating the main simulation loop (a single round trip).} The evolution of the field is then calculated in a split-step Fourier method \cite{Coen2012}, where dispersion and linear gain/loss are applied in frequency-domain, while saturation and the Kerr-effect are applied in the time-domain, where they are more naturally described.
We calculate the complex beam parameter for each moment in the fast time $t$, according to the instantaneous intensity at the same time.
The simulation is given a number of parameters - the nonlinear coefficient, the gain and loss, saturation parameter, pump mode size, slit size and aperture loss function, and then performed as follows:
\begin{enumerate}
	\item The field and complex beam parameter are initialized to random noise, simulating the seed spontaneous emission.
	\item For each round-trip, until the intracavity power stabilizes:
	\begin{enumerate}
		\item Calculate the instantaneous intensity $I_{n-1}(t)$ and the beam propagation matrix $M_{n}(t)$ for each moment in time. $M_n(t)$ depends on the instantaneous power through the Kerr-lens and changes in time.
		\item Propagate the complex beam parameter $q_{n-1}(t)$ by $M_{n}(t)$ to obtain the beam parameter $q_{n}(t)$.
		\item Calculate the instantaneous diffraction losses $l(w_{n}(t))$ and use that to calculate $I_{n}(t)$.
		\item Before the next steps, take the Fourier transform of the field $E_{n}(f)$ \textcolor{black}{and move to Frequency-domain, where linear propagation can be described as a simple transfer function}.
		\item Calculate the mean power over the roundtrip $\bar{I}_{n}$ and compute the gain saturation $g_{n}=1/(1+\bar{I_{n}}/I_{ss})$, \textcolor{black}{where $I_{ss}$ is the saturation power, calculated from the literature value and the steady-state beam waist.}
		\item Propagate the field through the cavity, including laser gain and dispersion \textcolor{black}{by applying it as a linear transfer function to the spectrum}.
	\end{enumerate}
	\item Return the temporal profile $I(t)$ and the beam parameter $q(t)$ as functions of time.
\end{enumerate}
\textcolor{black}{The loss function $l(w_{n}(t))$ can be any aperture function that is relatively flat and gives almost no loss within the boundaries of the aperture (taken to be the spatial size of the pump mode in the gain medium), and then rises rapidly as the beam diverges from this aperture.}
For the simulation we assumed a near threshold operation $g \approx l$, saturation power of $2.6 \rm{W}$ calculated based on the literature value and the stable beam size on the medium, a slit size of $ 38 \mu \rm{m}$ on the Kerr-medium, and dispersion-compensated operation.

\begin{figure}
	\centering
	\fbox{\includegraphics[width=0.85\linewidth]{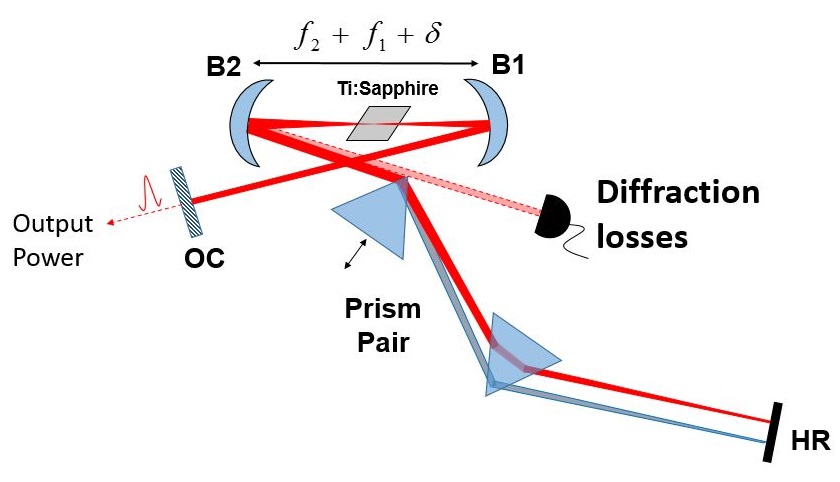}}
	\caption{Experimental layout. B1 and B2 are spherical mirrors with foci $f_{1(2)}$, HR is a highly-reflecting mirror, and OC is the output coupler. Light is measured both at the OC and near the first prism, where the diffracting light escapes. The penetration of the prism into the beam maybe varied to change the effective aperture size, as indicated by the arrow.}
	\label{fig:scheme}
\end{figure}

In our experiment, we measured the light leaking outside of the cavity, well separated from the main pulse as explained hereon. We employ a standard Ti:Sapphire oscillator in an X-folded four-mirror configuration \cite{yefet_2013}. Our gain medium is a 3mm Ti:Sapphire crystal located at a cavity focal point. The Ti:Sapphire crystal also provides the non-linearity for the Kerr-lens effect. 
Our cavity comprises of two spherical mirrors with focal length $f = 7.5\rm{cm}$, two planar end-mirrors, a prism pair (BK7 glass) and GDD mirrors to compensate for intra-cavity group delay dispersion (GDD) of $440 \rm{fs}^{2}$, and an output coupler with reflectivity of $R = 0.95$. The Ti:Sapphire has a \textcolor{black}{gain} bandwidth of $167 \rm{THz}$ and is pumped by a Verdi V18 $532 \rm{nm}$ laser, \textcolor{black}{with a much narrower oscillation bandwidth of roughly $20 \rm THz$}. The generated pulses have a repetition rate of $67\rm MHz$ and a duration of roughly \textcolor{black}{$50 \rm fs$.}

\begin{figure}[ht!]
	\centering
	\fbox{\includegraphics[width=0.9\linewidth]{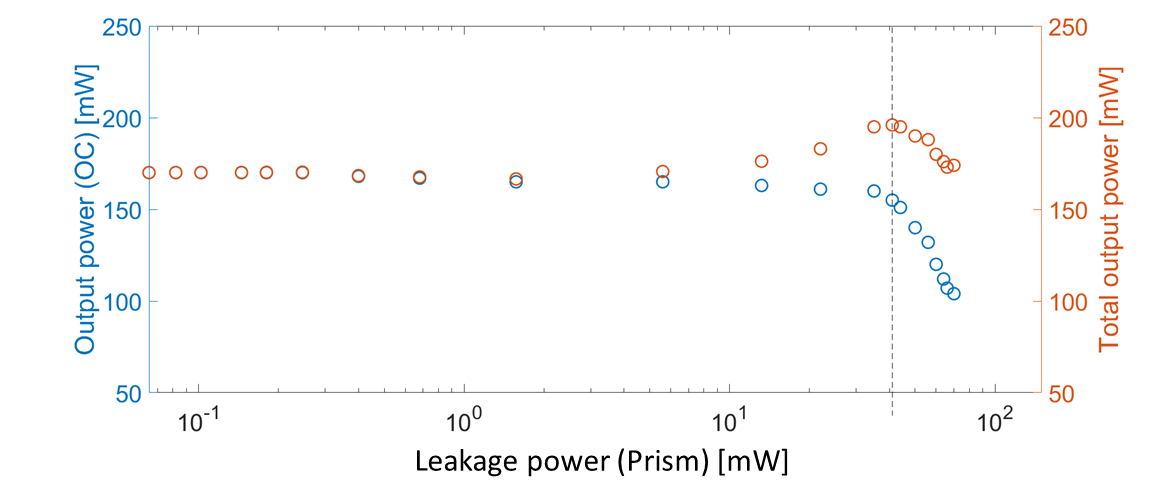}}
	\caption{Main pulse power at the output coupler vs. the leakage power at the first prism. Total power \textbf{(orange)} and output power \textbf{(blue)} as function of leakage power. The dashed line indicates the maximal leakage power diffracted out due to the Kerr-lens effect.}
	\label{fig:power}
\end{figure}

Since the laser beam passes near the edge of the prism pair, it also acts as an effective slit for the beam that allows to measure the light that leaks out of the cavity due to diffraction, as shown in Fig. \ref{fig:scheme}. However, we stress that the prism pair only acts to isolate the diffracted light, and does not add any losses on its own. The leaking light is not lost due to the prism, but due to the temporal Kerr lens variations. The prism only allows us to observe and isolate the diffracted light, thus changing the losses at the prism does not directly affect the pulse duration and power of the main output. This is indeed observed on the spectrum and output power that do not depend on the prism position, as shown in Fig \ref{fig:power}.
The distance between the mirrors is offset by $\delta$ from a perfect telescope, corresponding to different stable (or unstable) ray configurations, with our cavity operated just beyond the cavity stability zone, which induces substantial diffraction losses for CW operation, as illustrated in Fig. \ref{fig:configuration}. We note that operation at the edge of the stability zone is common, but not strictly required, and KLM had been achieved in stable configurations as well \cite{Modsching2019}. Beyond the stability zone, continuous-wave (CW) operation is unstable and suffers diffraction losses and the cavity stability becomes extremely susceptible to changes in the optical power inside the cavity due to the Kerr-lens effect. Consequently, the Kerr-lens counteracts the diffraction losses and pushes the cavity back into stable operation, leading to differences in the loss between high and low intensity operation, inducing the saturable diffraction losses that act as an effective fast saturable absorber. \textcolor{black}{The strength of the effective saturable absorber can be controlled by adjusting moving further outside the stability zone, and changing the stable waist size.}
Clearly, KLM is a spatio-temporal effect, i.e. the instantaneous spatial and temporal profiles are closely linked and directly affect one another. Explicitly in our configuration, when the instantaneous intensity is high, the non-linear lens is strong, resulting in a smaller beam waist radius on the prisms and no leakage out of the cavity. This allows for effective isolation of the diffraction losses from the main pulse. The dynamics of the Kerr-lens is clearly captured by the spatio-temporal profile of the leakage pulse, shown in Fig. \ref{fig:temporal_profile}.

At peak intensity the beam size is small and barely any light leaks out of the prism, whereas at the leading and trailing edges of the pulse, it is large and a fraction of the light leaks out of the cavity, but since the overall power is rapidly decreasing, not much is leaked from the cavity at the pulse edges. The major amount of power leaks between the main peak and the edges, where there are substantial diffraction losses that still carry power and measurable power leaks outside. Consequently, the light that leaks outside of the cavity should have the shape of a double pulse, with two peaks centered around the main pulse, as shown in Fig. \ref{fig:autocorrelation}. The temporal profile of the leaking power depends on the instantaneous pulse power, and the shape of the aperture $P_{leak}(t) = \int dx P(t) \cdot f(x)$, where $f(x)$ is the aperture function, \textcolor{black}{which in our experiment is implemented by the prisms} and modeled as a slit in our simulation. The exact shape of the aperture does not change the properties of the leaking light, but its size directly affects the temporal shape of the pulse and the time delay between the peaks, smoothly transitioning from a single peak to a double peak. \textcolor{black}{This does not affect the actual profile of the pulse inside the cavity, but rather only the temporal shape of the coupled light. It does not affect the actual diffraction losses, which are due to the virtual aperture induced on the gain medium by the KLM effect.}


In the experimental scheme described above, we measured the light leaking from the prisms in both frequency (spectrum) and time (intensity auto-correlation), and compared them to the same measurement of the main pulse out of the output coupler, as shown in figure \ref{fig:autocorrelation}. The time scales and qualitative features of the measured and simulated pulse agree very well. Indeed, the intensity auto-correlation shows a triple-peaked profile, indicative of a double pulse, as predicted by our theoretical analysis, and as discussed above. The spectrum of the leaking light is broader than that of the main pulse, since it is shorter temporally.
\textcolor{black}{There is some discrepancy between the theoretical and experimental auto-correlation profiles - the temporal separation of the side lobes of the autocorrelation from the main peak is $\sim90\rm{fs}$ in the experiment, but the theoretical separation is $\sim60\rm{fs}$. The agreement between the numerical simulation and the experiment should not be assumed to be fully quantitative, since we make a number of simplifying assumptions, such as thin Kerr-lens approximation and a perfect Gaussian spatial mode, that capture the main important dynamics of Kerr-lens mode-locking in a simple way, but miss exact quantitative agreement. In addition, we note that the experimental autocorrelation is subject to imperfections, which may also cause discrepancy from the theoretical results.} This is due to the slow response at high gain of the auto-correlator's photo-detector compared to the ultrafast pulses. The pulses generated by our oscillator were barely long enough to measure on our auto-correlator, which resulted in a distorted, asymmetric shape, which also leads to a ``smearing'' of the auto-correlation curve in the direction of the delay scan. The contrast ratio between the main and satellite peaks, which is less than 0.5, may indicate that the laser pulse in the experiment was slightly asymmetric in time (as opposed to the simulation that assumed a perfectly symmetric pulse). Such asymmetry would carry on to the leaking pulses, as indicated by the asymmetric measured spectrum of the leaking pulse (see Fig. \ref{fig:autocorrelation}). Regardless, the important properties of the auto-correlation profile are clear, and the spectrum of the leaking pulse matches that of a double pulse.
The double pulse shape is only observed for the light leaking at the prism, and not for the light coupled outside in the output coupler, which gives a strong indication that the double pulse observed on the leaking light is indeed due to the Kerr-induced diffraction, and not other effects.

We further confirm that the power leakage we observe is not just standard linear loss of the main pulse, but indeed power leaking due to the saturable diffraction losses, by observing both the leakage power and the output power as the prism is shifted further in/out of the beam (see \ref{fig:scheme}). We observe that moving the prism varies the effective slit size, letting more/less of the diffracted light to leak out of the cavity, but that the main pulse power is hardly affected by that. Figure \ref{fig:power} shows the the main output power as a function of the leakage power, indicating clearly that the power of the main pulse is independent from the leaking light (up to the physical limit of clipping the main beam). 

In conclusion, we directly observed the diffraction losses in a KLM mode-locked laser in the form of the light leaking out a dispersion compensating prism pair. The temporal intensity auto-correlation of the leaking light is intimately related to the spatio-temporal profile of the pulse, and reveals important properties of the light leaking due to diffraction. Our experimental apparatus provides very good isolation between the main pulse and the diffraction losses, and the observed experimental results match our analysis. We show that the leaking light and the main pulse are distinct, and have very different temporal intensity auto-correlation profiles. The ideas and methods we presented can be developed as a tool to study the spatio-temporal dynamics of lasers, and open a new window towards studying the ultrafast spatial dynamics of Kerr-lens mode-locked lasers in particular and spatio-temporal mode-locking in general. 

\bibliography{prism_bib}



\end{document}